# COLLABORATIVE KNOWLEDGE CREATION AND MANAGEMENT IN INFORMATION RETRIEVAL

Victor Odumuyiwa[1]
Victor.odumuyiwa@loria.fr
Amos David[1]
Amos.david@loria.fr
[1]Equipe SITE-LORIA, Nancy Université
Laboratoire Lorrain de Recherche en Informatique et ses Applications
Campus Scientifique – BP 239 54506 Vandœuvre lès Nancy Cedex, France.

Abstract: The final goal of Information Retrieval (IR) is knowledge production. However, it has been argued that knowledge production is not an individual effort but a collaborative effort. Collaboration in information retrieval is geared towards knowledge sharing and creation of new knowledge by users. This paper discusses Collaborative Information Retrieval (CIR) and how it culminates to knowledge creation. It explains how created knowledge is organized and structured. It describes a functional architecture for the development of a CIR prototype called MECOCIR. Some of the features of the prototype are presented as well as how they facilitate collaborative knowledge exploitation. Knowledge creation is explained through the knowledge conversion/transformation processes proposed by Nonaka and CIR activities that facilitate these processes are highlighted and discussed.

## 1. Introduction

Observations of users' information behaviour show that users collaborate during information seeking and retrieval processes. We observed that users in real world manifest Collaborative Information Behaviour (CIB) whereas most of the existing information systems model users without taking note of their collaborative behaviour.

Such behaviours manifested by users in resolving their information problems include:
- consulting multiple information sources and other people in solving their information problem
- monitoring the evolution of an information system that they consider as a source of relevant information
- depending on their social and professional network in solving their information problem
- exploiting the network of experts in their domain in order to solve their information problem
- giving more credit to information given by an expert than information retrieved from an Information Retrieval System (IRS)

These behaviours confirm submissions and arguments of IR researchers that IR is a social and cognitive process (Karamuftuoglu 1998, Wilson 1981). Many authors concurred to the fact that the global objective in information retrieval is the creation of knowledge (Robert 1999, Karamuftuoglu 1998). David and Thierry (2002) adopted the evocative habits in human learning to support the cognitive aspect of information retrieval and explained how these evocative habits should influence the design of information retrieval system. They highlighted four evocative habits of a learner (in our case an IRS user):
- Observation process: this process is the discovery process i.e. it allows a user to discover the domain of study.
- Knowledge acquisition process: this is the process through which a user uses its already acquired knowledge to acquire new knowledge.
- Knowledge application process: this is the application of already acquired knowledge for problem solving.
- Creativity process: from the acquired knowledge, a user can create a new knowledge which is unique to him. This process varies for each user and it is at this stage that a user materializes the experience gained from the domain.

From these evocative habits, they developed a functional IR model comprising of four cognitive tasks in IR: exploration, query, analysis and annotation (David & Thiery 2002). These evocative habits and the ensuing functional IR model support knowledge acquisition, sharing and creation in IR.

The importance of knowledge sharing or learning in IR is also seen in all the new technologies where users' models are being integrated in IRS development. The growing rate of collaborative tagging (Golder and Huberman, 2006) confirms the sharing culture which has come to stay in IR. Collaborative bookmarking is also in the increase and many users are benefiting from the collective intelligence of community of users. All these point to the importance of creating and enhancing knowledge through collaborative effort in information retrieval. The inclusion of social network in information retrieval, given rise to social information retrieval (Evans & Chi 2008, Kirsch 2005), is also improving the way information problems are getting solved. Even though this collective intelligence appears to better the lots of users in satisfying their information needs, it is of paramount importance to note that these approaches are more of cooperation than collaboration because users do not share a common goal or objective even though they share their knowledge (Odumuyiwa & David 2009). In CIR, we are interested in a situation where users share the same information problem and based on this, go ahead to share their knowledge and create new knowledge in solving the problem. This leads us to consider the following research questions:
- How is knowledge created in CIR?
- How do we organise or structure created knowledge in CIR?
- How do we exploit created knowledge?

We shall at first explain the concept of CIR in section 2. The process of knowledge creation and conversion in CIR will be treated in section 3. In section 4, our focus will be on organising and structuring created knowledge in CIR. Section 5 will be on the functional architecture of our CIR prototype and how it facilitates knowledge exploitation. Finaly, we conclude in section 6.

## 2. Collaborative information retrieval

We see CIR to be at the intersection of information search, communication, knowledge management and social networks. This corresponds partly with the report of the first collaborative information retrieval workshop (Pickens et al, 2008) wherein CIR was considered to be at the intersection of communication and search which are the two most common uses of Internet technologies. To us CIR entails more than search and communication because new knowledge is produced during collaboration hence such knowledge should be managed (organised, structured, and stored) for reuse and more so for collaboration to take place, there must be a bringing together of collaborators or rather put a networked community of users.

In order to understand CIR, there will be need to define the two underlying concepts - collaboration and information retrieval. According to Fidel et al. (2006), information retrieval can be interpreted in a broader sense to include processes such as problem identification, analysis of information need, query formulation, retrieval interactions, evaluation, presentation of results, and applying results to solve an information problem. Collaboration can be defined as the act of working jointly with shared objective. It is important to note that there is a difference between collaboration and cooperation. In collaboration, users mutually share their knowledge towards a shared goal while in cooperation users may not necessarily have a common goal (Longchamp, 2003). They can share their knowledge but may not necessarily have a shared goal.

Golovchinsky et al (2008), in trying to analyze and classify systems that support collaborative information seeking, proposed a model which contains four dimensions: intent, depth, concurrency and location. Using the "intent" dimension of their model, they explained what collaboration in IR should be. The intent can either be implicit and explicit. An example of an implicit collaboration exists in recommender systems in which the behaviour of a group of users with respect to a particular information object is used to suggest choices to others searching for similar information. "While people may be generally aware that their results are based in part on data obtained from other users, they may not know who those people are and what purpose they had in mind while searching. ...In some sense, this is not strictly collaboration but rather a coordination of people's activities." (Golovchinsky et al, 2008). Explicit collaboration on the contrary entails a group of users searching for documents to meet a **shared** information need.

The "depth" dimension explains the type of mediation supported in information seeking. While some systems use UI (User Interface)-only mediation, others use deeper algorithmic mediation. The "concurrency" dimension differentiates between systems that support synchronous collaboration and those that support asynchronous collaboration. The "location" dimension distinguishes co-located collaboration from distributed collaboration.

We are interested in a synchronous explicit CIR with both UI and algorithmic mediation that can support collocated and distributed collaboration. Hence we consider CIR as consisting of methods and systems for managing collective activities of users in information retrieval process in order to facilitate direct collaboration among the users thereby enabling knowledge sharing among them. The emphasis here is not only on the "collective activities" but also on the direct collaboration among the users which leads to sharing of tacit knowledge.

## 3. Knowledge creation in CIR

Knowledge creation has been recognized to be strategically important for organizational learning and innovation. Many of the works on knowledge creation are centred on organization learning with the goal of facilitating knowledge conversion and transformation from and within its two forms: tacit and explicit knowledge (Nonaka, 1991; Nonaka and Takeuchi, 1994). Tacit knowledge can be defined to be the "knows" and the "know-hows" which are manifested in actions and activities. Marwick (2001) put it as what knower knows which is derived from experience and embodies beliefs and values. Explicit knowledge is articulated or expressed knowledge represented by artefacts such as written document, electronic document, video etc. They are easily transferable and were created with the goal of communicating with another person.

Nonaka (1991) highlighted the processes by which knowledge is transformed within and between forms usable by people. These processes include: socialization, externalization, combination and internalization.
- Socialization supports tacit to tacit knowledge transformation. People acquire tacit knowledge through their interaction and communication with others especially in team meeting where experiences are described and discussed.
- Externalization supports tacit to explicit knowledge conversion. This process allows capturing users' tacit knowledge in explicit form through conceptualization, elicitation and articulation during collaboration with others (Marwick, 2001).
- Combination is the process of transforming explicit knowledge to explicit knowledge. An example is the classification of a document or the addition of metadata to a document. The document itself is an explicit knowledge while the metadata is a new explicit knowledge created on the document.
- Internalization is the process of transforming explicit knowledge to tacit knowledge. This is the process by which individuals internalize information so as to create their own tacit knowledge.

Marwick (2001) highlighted some technologies that may be applied to facilitate these knowledge conversion processes with emphasis on organisational context. All these four knowledge conversion process also applies to collaborative infor-mation retrieval. We highlight in table 1 some CIR activities through which knowledge transformation takes place in CIR.

Table 1: CIR activities that facilitate knowledge conversion processes

| Tacit to Tacit (socialization) | Tacit to Explicit (externalization) |
|---|---|
| Interpersonal communication | Problem definition and clarification |
| Synchronous IRS interface sharing | Annotation |
| **Explicit to Tacit (internalization)** | **Explicit to Explicit (combination)** |
| Visualization of search history | Tagging, Query formulation, Metadata creation, Classification |
| Consultation of search results | |

*3.1 Socialization in CIR*

Interpersonal communication allows users to engage in synchronous interaction during information retrieval through audio, video or textual information exchange. This facilitates experience sharing, hence culminates in acquisition of tacit knowledge. Synchronous IRS interface sharing makes it possible for a user to observe another user's activities through the WYSIWIS[1] technology. It is a way of communicating one's competence in real time while solving information problem.

*3.2 Externalization in CIR*

Information problem definition is the articulation or elicitation of information problem to be solved. This definition is a factor of user's domain knowledge. The tacit knowledge of a user is made explicit during information problem elicitation. Annotation facilitates knowledge creation in IR. A user's annotation is seen as an added value to information. Figure 1 shows the information problem annotation interface of our prototype which allows users to clarify their information problem.

*3.3 Combination in CIR*

When a user formulates a query, the query is a representation of the user's conception of his information problem as well as his conception of the functioning of the information retrieval system being used for search. Hence, a user's query can be considered as an explicit knowledge on a defined information problem. When the query is formulated collaboratively, it can be considered as an explicit knowledge on a shared information problem. Tagging and metadata creation are also knowledge creation activities through which explicit knowledge are created from explicit knowledge such as documents, videos etc. Document classification and indexing can be regarded as a form of explicit to explicit knowledge conversion (Roberts, 1990). When a user or a domain expert analyses a document and then add a metadata to categorize or classify the document, the metadata is an enhancement of the explicit knowledge artefact i.e. the document.

---

[1] WYSIWIS: What you see is what I see.

*3.4 Internalization in CIR*

Visualization of search history enables tacit knowledge acquisition. A user visualizing the past search activities of others would learn from the experience of others. Consultation of search results is also an internalization process. A user reading a document retrieved during IR process, internalizes the information contained in the document to acquire tacit knowledge which he may later externalize through annotation to produce an explicit knowledge.

Figure 1. Information problem annotation interface

## 4. Organizing created knowledge in CIR

Many works have been done on facilitating the activities through which internalization and combination knowledge conversion processes occur in CIR (Golder and Huberman, 2006; Zhang and Li, 2005). Our major interest in this paper is on the activities that facilitate the sharing and conversion of tacit knowledge through the socialization and externalization processes in collaborative information seeking and retrieval.

*4.1 Problem definition*

The first activity is the problem definition and clarification between the collaborators. This is the first phase in CIR and it determines how successful the collaboration activity would be. This phase allows the collaborators to integrate and differentiate their understanding of the information problem at hand in order to arrive at a shared understanding of the problem. To organize and structure the knowledge created during problem definition, we use the following attributes: information problem, objective, date/time stamp, domain, keywords, information sources and indicators.

Information problem is what necessitates the search for information i.e. the information need of the user. At the initial stage in IR, this can be vague and may need to be analyzed. For example, a user seeking for information on cassava plantation in West Africa has a broad information problem which needs to be analyzed. But to be able to analyze this problem, another information is needed which is the search objective. An information problem may have to be decomposed into smaller information problems during the clarification process. Hence the collaborators may have to define some attributes and values explicitly during the clarification process. This leads to dynamic creation of tables in the database on the fly by users during collaboration process.

Objective is the "why" of the information search. It explains the goal of the search and dictates the direction of search. During collaboration, the objective may not change but it may lead to change in the information problem. The information problem will have to be interpreted, analyzed and clarified based on the search objective.

Date/time stamp is to contextualize the problem definition in case of evolution in the definition as the collaborators' domain knowledge increases. It is automatically added by the system each time a modification is made on the problem definition.

Domain: users can attribute their information problem to a domain. This helps to reduce the problem of interpretation of concepts that cross domain with different interpretation in each domain. It also helps in the eventual retrieval (for reuse) of the knowledge captured during the collaboration.

Keywords: users can suggest keywords to describe the information problem at hand. These keywords are to be used to eventually formulate queries in the course of search.

Information source: part of the problem definition clarification is the identification of likely information sources to be consulted in the search process. Information sources can include systems and people.

Indicator: users can define indicators to specify what they think must be present in retrieved documents in order to judge them relevant. The indicators are defined as attribute-value pair.

*4.2 Problem clarification*

Taking the example of the information problem given above, the information problem is on cassava plantation in West Africa. The objective is to know from which company to import cassava from. The collaborators exchange their views and perceptions about the problem. For example, a user suggests looking for information on the major exporters of cassava in West Africa. Another may suggest looking for countries in West Africa that produce cassava and the regions engaged in the production as well as problems related to production in such areas. Another user may suggest looking at the major importers of cassava from West Africa. Another user may suggest looking at the various products that are made from cassava and the companies involved in the production of such products because this can help in determining the demand rate for cassava in the market. Other suggestions could be: qualities of cassava production in West Africa; cassava exporting agents in West Africa; local consumption of cassava in Nigeria which is the world largest producer of cassava[2].

This decomposition is done collaboratively and it depicts the level of domain knowledge of the collaborators. We could see that an information problem that a user might lightly put on Google as strings of text without actually getting a satisfying response has been broken into sub-information problem just because the objective of search is indicated. It is also important to note that this clarification which involves interpersonal communication among the collaborators cannot be totally automated. It can only be handled by mediating the interaction between collaborators. All the sub-information problems collaboratively agreed upon would be added through annotation to the original information problem.

The next step in the clarification would be to suggest relevant information sources to use for search. Naturally, most users will suggest Yahoo, Google etc. but other suggestions such as the website of the International Institute of Tropical Agriculture (IITA), open archive on agriculture (www.e-agriculture.org) etc. could also be made. This leads to suggestion of likely keywords for search. This is followed by attribution of the problem to domain. The above problem may be related to agriculture, nutrition, international business, exportation and importation.

Even though searching or retrieval has not yet started, the above attributes have been used to structure the knowledge expressed by the collaborators during the problem definition and clarification process. After the collaborators have arrived at a shared understanding of the information problem, they can start formulating query based on the elements defined in the problem definition phase. They keep engaging in interpersonal communication and sharing of search results during this phase. Their queries, the documents retrieved and the evaluations of the retrieved documents are shared among them and they are also captured and capitalized for future reuse.

Figure 2 shows a screen copy of the main interface of a prototype, developed in java, wherein we implemented our approach. The system allows users to communicate synchronously during collaboration. It also allows switching between problem definition interface and the online search interface. Users can consult any online information source through the system. All these functionalities are integrated into a single environment which differentiates it from existing approaches. As the users are suggesting information sources and likely keywords through the problem definition interface, they could at the same time consult suggested information sources and try out queries based on suggested keywords.

---

[2] http://www.iita.org/cms/details/cassava_project_details.aspx?zoneid=63&articleid=267

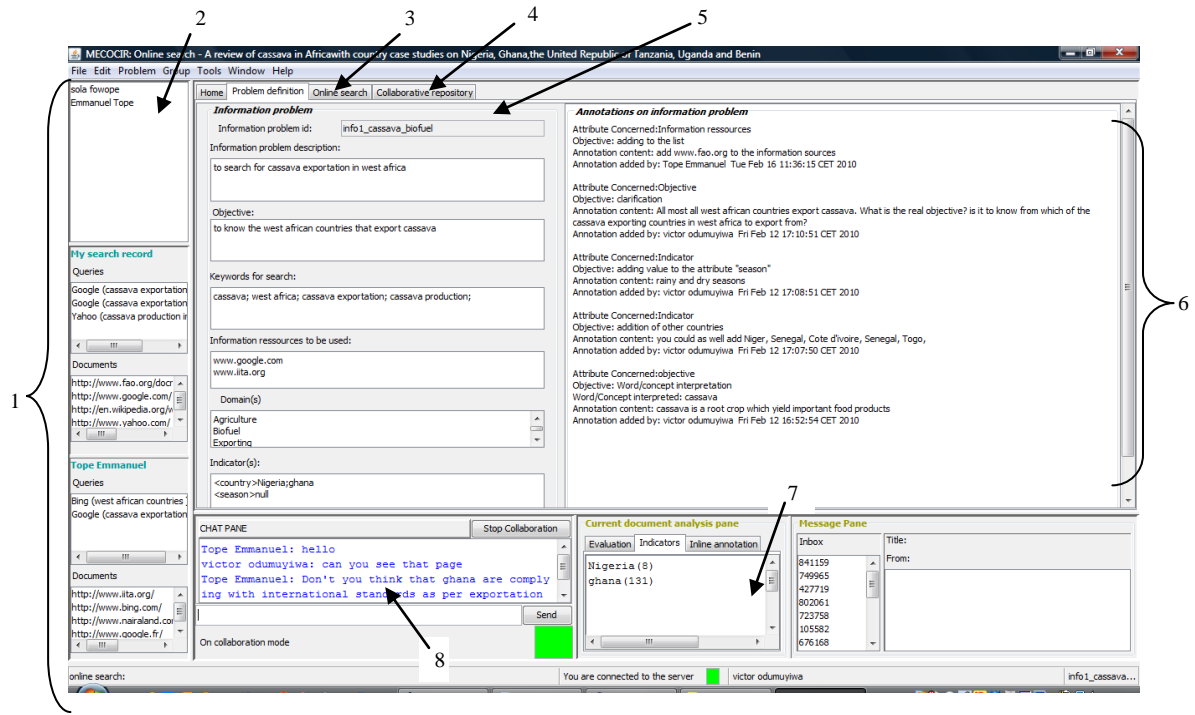

Figure 2. MECOCIR main interface. (1) Awareness interface, (2)Collaborating users list, (3) Browser interface tab for online search, (4) Collaborative repository interface, (5) Information problem definition interface, (6) Annotations on information problem (7) Indicator analyzer output (8) Instant messaging

# 5. Exploiting created knowledge in CIR

In the previous sections, some details had already been given concerning our prototype. This notwithstanding, we shall explain here below the functional architecture of our prototype and how this facilitates knowledge exploitation in CIR.There are three main layers present in our functional architecture as shown in figure 3. These include: the CIR users' interface layer, the CIR system core layer and the data/knowledge layer.

## 5.1 CIR User interface layer

The CIR users' interface layer which serves as the collaborative workspace for the CIR system users is made up of five other interfaces: browser interface, problem definition interface, awareness interface, collaborative repository search interface, and annotation interface.The browser interface permits users to connect to any online information sources including search engines and dedicated databases. The problem definition interface allows a user to define and clarify his/her problem as explained earlier under the knowledge organisation section. The awareness interface allows collaborating users to know in real time what their partners are doing. It shows those that are online. It also displays all the queries formulated by them and the search engine to which they submitted the queries. It shows all the documents visited by one's partner as well as the current document being consulted by him/her. In fact this awareness scheme facilitates synchronous knowledge sharing among users. A user can click on the current document being viewed by his partner which in turn will be displayed on his/her screen.

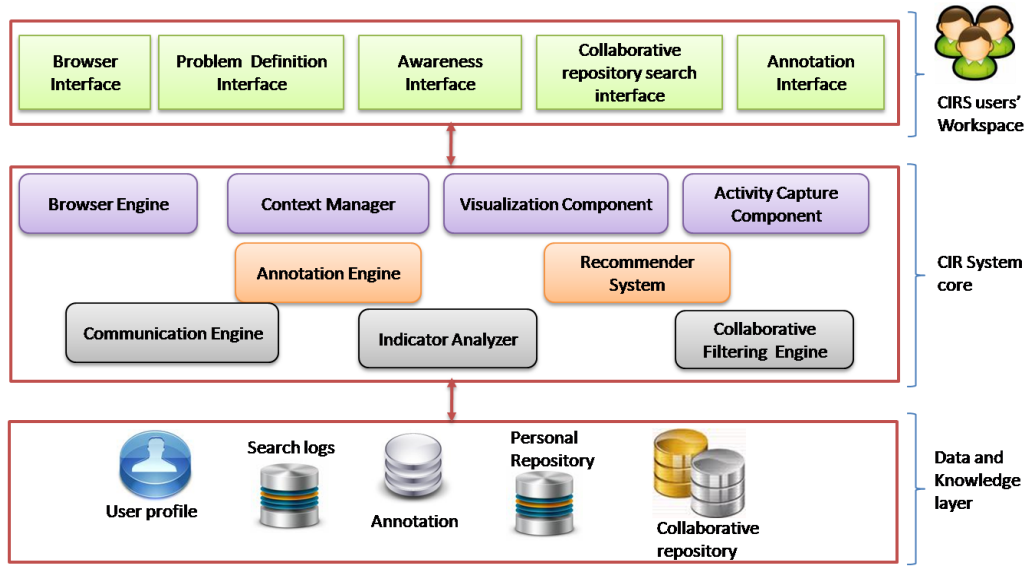

Figure 3. Functional architecture of MECOCIR collaborative information retrieval environment

The collaborative repository search interface allows users to search for solution from the gathered collective intelligence. It allows a form of implicit collaboration and may be a starting point to getting solution to a problem. This interface also furnishes the user with system's recommendation of potential collaborators by comparing the user's problem with past captured and stored problems on the collaborative repository. This is achieved by calculating the similarity between the current problem and the past problems. Users that participated in the resolution of past similar problems are then suggested as potential collaborators to current user. This can lead to a demand for explicit collaboration from such users if they are online. The annotation interface as explained earlier allows users to create value added information. This interface allows for evaluation of retrieved documents with regards to the information problem at hand. It also allows adding clarification to an information problem as shown in figure 1.

### 5.1 CIR System core layer

The core of the CIR environment contains eight major components: the browser engine, context manager, visualization component, activity capture component, annotation engine, recommender system, communication engine, indicator analyzer and collaborative filtering engine. Most of these components serve as engines for the interfaces hence they perform the operations which are displayed on the interfaces. The browser engine and activity capture component handles back-

 processing for the browser interface. Back-end processing for collaborative repository search interface is handled by the collaborative filtering engine and the recommender system. The communication engine and context manager handles processing for the problem definition interface and awareness interface while the annotation engine feeds the annotation interface. The indicator analyzer analyses each document being viewed by the user thereby giving the frequency of occurrence of the user's defined indicators in each documents. This helps the user in judging the document relevance to the information problem at hand. For the moment this analyzer performs statistical analysis. Our aim is to improve this by upgrading it to a semantic analyzer so as to better capture the semantic elements of the attributes and values representation of indicators.

*5.2 Data and Knowledge layer*

The data and knowledge layer consists of the various database and knowledge base used for storing captured knowledge and data.

# 6. Conclusion and future work

In this paper we considered knowledge creation and management through collaborative information retrieval. We supported the assertion that information retrieval is a social as well as a cognitive process having knowledge production as the final goal. Our major focus was on the exploitation of tacit knowledge of collaborators during CIR to create explicit knowledge and to facilitate tacit knowledge acquisition. This was exemplified in the problem definition and clarification phase through interpersonal communication and annotation creation centred on an articulated information problem. We carried out the first series of experimentation by allowing users to collaborate in solving real life information problem using our prototype. We paired the users and gave them an information problem to solve. To our surprise, we discovered that all the users involved in the experimentation have different approaches to solving their information problem. In the problem definition and clarification phase, a great level of complementarities in peers' knowledge was observed and much new knowledge was created and shared between the pair. This knowledge was captured and stored for future reuse. Users that participated in the experimentation expressed how the prototype helped them to share their knowledge. We can conclude that the CIR activities listed in table 1 facilitate knowledge creation in CIR and that our prototype aids in managing and exploiting created knowledge in CIR. Our major perspective is to carry out a more elaborate experiment involving many users in order to draw more reasonable conclusions.